\documentclass[
final,
5p,
times,
twocolumn,
authoryear,
amssymb
]{elsarticle}

\usepackage{bm}
\usepackage{dcolumn}% Align table columns on decimal point
\usepackage{graphicx}
\usepackage[utf8]{inputenc}
\usepackage[english]{babel}
\usepackage{amssymb}
\usepackage{amsmath}
\usepackage{amssymb}
\usepackage{mathtools, cuted}

\newenvironment{equationc}{\begin{equation}}{,\end{equation}\ignorespacesafterend}
\newenvironment{equationp}{\begin{equation}}{.\end{equation}\ignorespacesafterend}

\begin{document}

\begin{frontmatter}

\title{Revisiting the 3$\alpha$ reaction rates in helium burning stars}

%%----------------------------------------------------------------------------------
%%%
%%% Author list in elsart style:
%%%

\author{ T. Depastas$^{a,*}$, S.T. Sun$^{b}$, H.B. He$^{b}$, H. Zheng$^{b,*}$ and A. Bonasera$^{a,c}$ }

%--------------------------------------------------------------
%--------------------------------------------------------------

\address{ $^{a}$ Cyclotron Institute, Texas A\&M University,
                     College Station, Texas, USA }
\address{ $^{b}$ School of Physics and Information Technology, Shaanxi Normal University, Xi’an 710119, China }
\address{ $^{c}$ Laboratori Nazionali del Sud, INFN, Catania 95123, Italy }

\address{ $^{*}$ Corresponding authors. Emails:  tdepastas@tamu.edu, zhengh@snnu.edu.cn}
%%************************************************************************
%abstract
%*******************************************************************

\begin{abstract}
Helium burning is one of the most fundamental steps of stellar nucleosynthesis, as it describes the formation of life-determining element of carbon, while it plays a key role in the evolution of Red Giant, accreting White Dwarfs and Neutron Stars. In this work we develop a generalized statistical theory for the 3$\alpha$ reaction, which is based on the use of the Imaginary Time Method, along with the semi-classical Hybrid $\alpha$-Clustering (H$\alpha$C) and Neck Model (NM) frameworks. The results compared to the methodology and data of the NACRE collaboration, following in several orders of magnitude discrepancies, especially at low temperatures. This may be crucial for the early dynamics of helium burning stars.
\end{abstract}

\begin{keyword}
%% keywords here, in the form: keyword \sep keyword, up to a maximum of 6 keywords
Triple alpha reaction \sep Sub-barrier fusion \sep Imaginary Time Method \sep Helium Burning \sep Stellar Reaction Rates

%% PACS codes here, in the form: \PACS code \sep code

%% MSC codes here, in the form: \MSC code \sep code
%% or \MSC[2008] code \sep code (2000 is the default)

\end{keyword}

\end{frontmatter}

\begin{figure}[!ht]                                        %%% htbp
\hspace*{-0.5 cm}
\centering
\includegraphics[height=8.5 cm]
{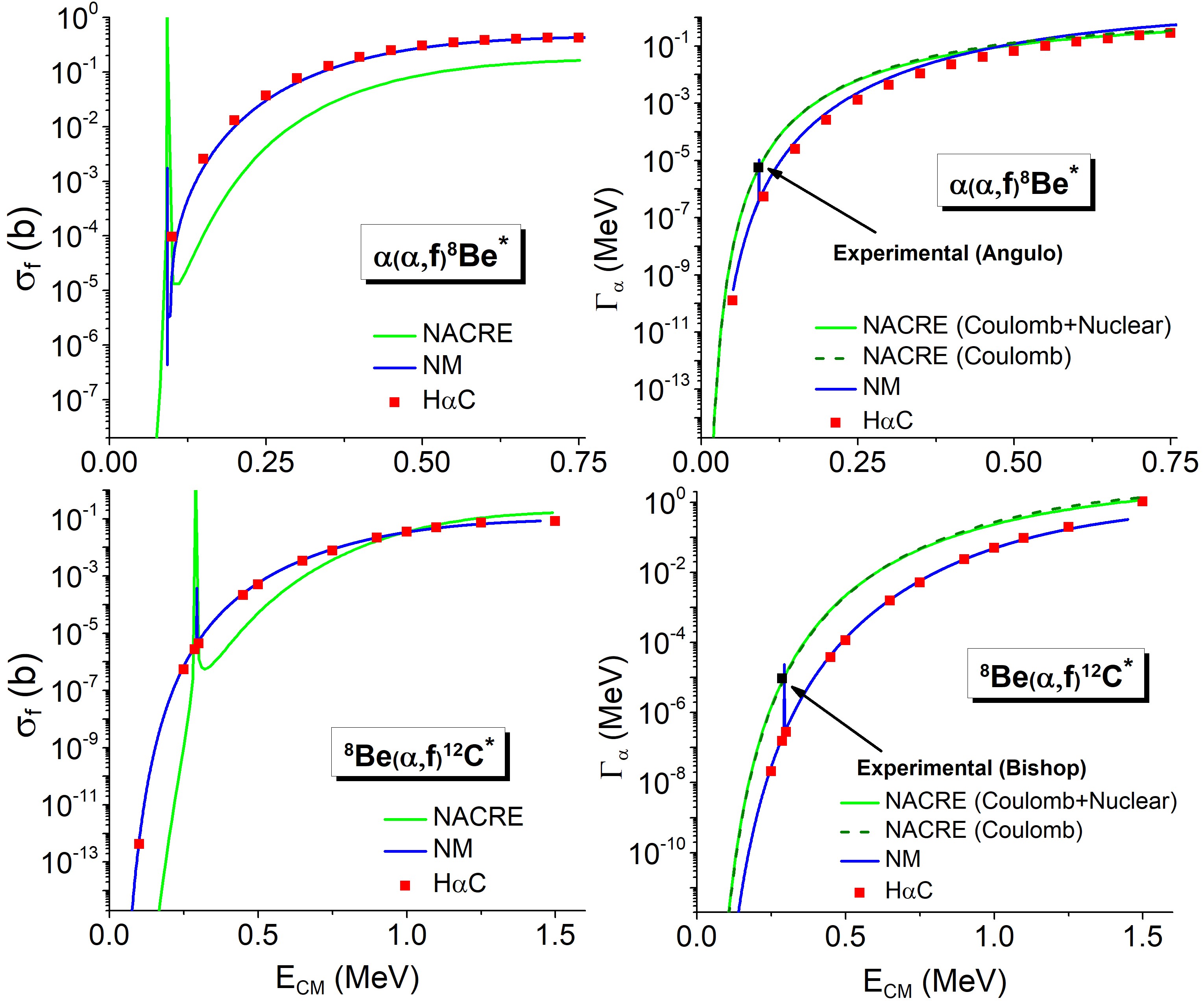} 
\caption{(Color online) Fusion cross sections (left column) and $\alpha$ widths (right column) for the $\alpha(\alpha,f){}^{8}\textit{Be}^*$ (top) and ${}^{8}\textit{Be}(\alpha,f){}^{12}\textit{C}^*$ (bottom). The results of NACRE (\cite{Angulo1999}), H$\alpha$C and NM are given according to the key. The experimental widths (\cite{Angulo1999,BishopEfimov2021}) are noted by black points.}
\label{Fig1}
\end{figure}

\begin{figure}[!ht]                                        %%% htbp
\hspace*{0.0cm}
\centering
\includegraphics[width=7.5 cm]
{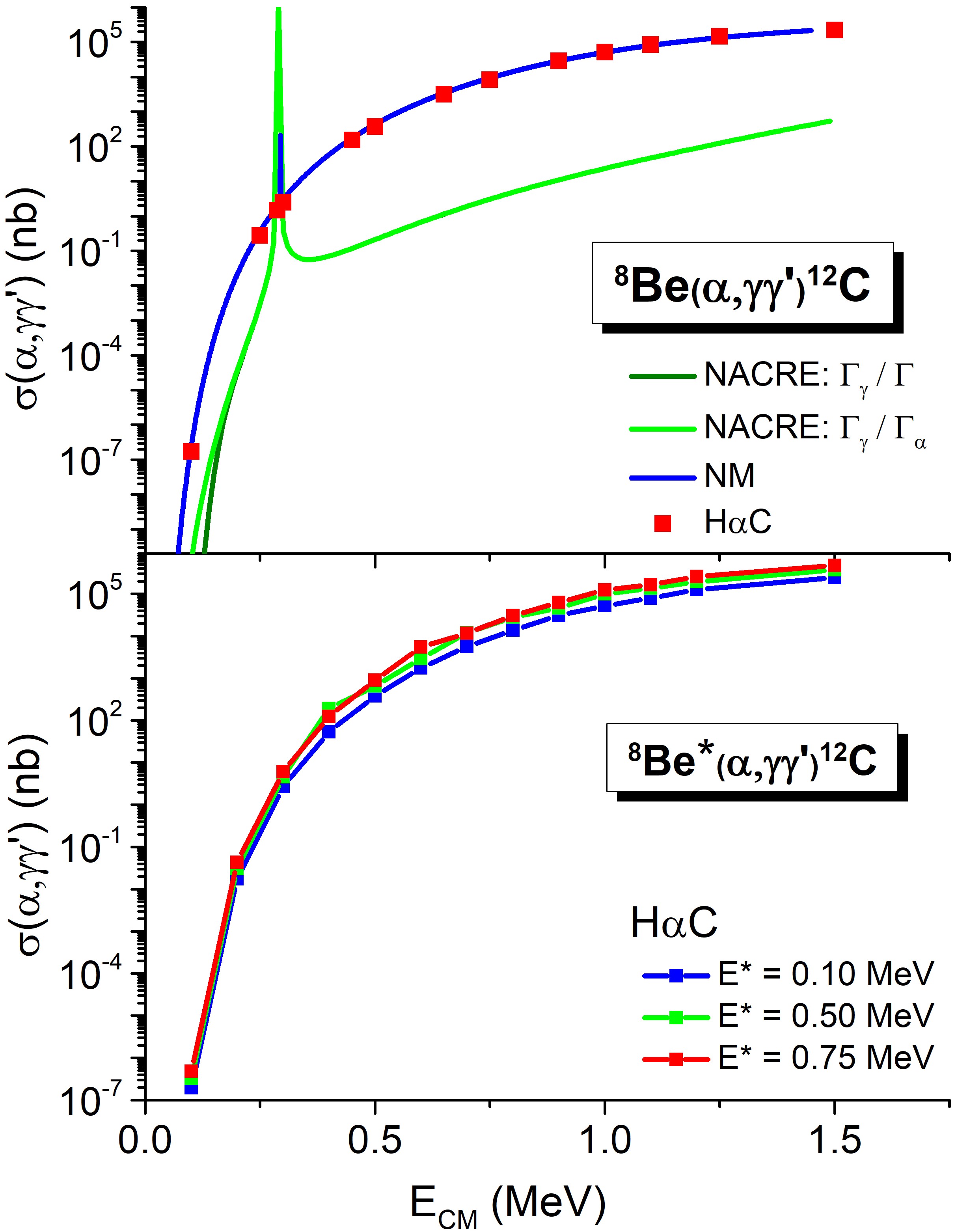} 
\caption{(Color online) The cross section for the ${}^{8}\textit{Be}(\alpha,\gamma\gamma'){}^{12}\textit{C}$ reaction, with the beryllium in its ground state (top) and in the excited continuum (bottom). The results of NACRE in its original form (\cite{Angulo1999}) and with the width correction, as well as the data from the H$\alpha$C and NM calculations, are given according to the key.}
\label{Fig2}
\end{figure}

\begin{figure}[!ht]                                        %%% htbp
\centering
\includegraphics[width=7.9 cm]
{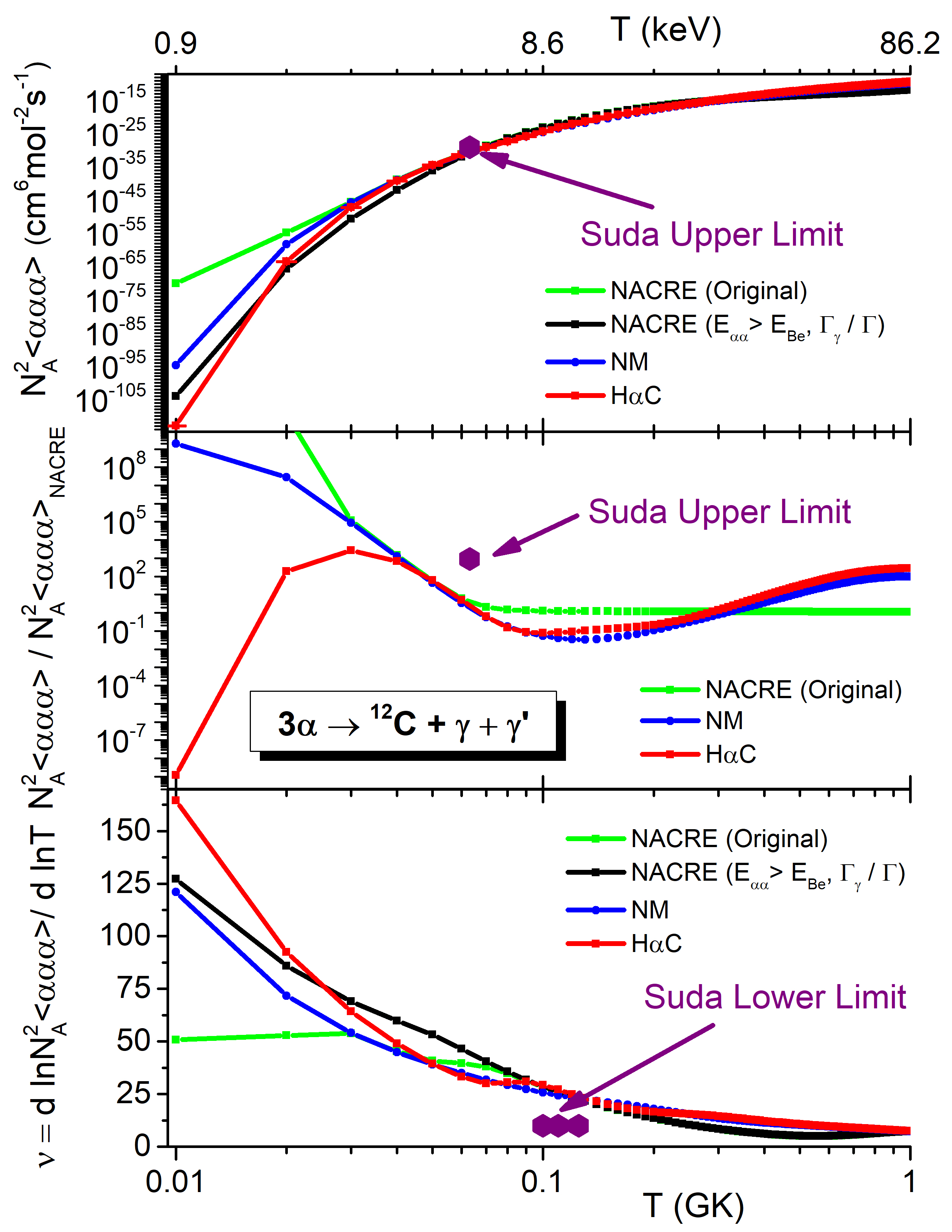} 
\caption{(Color online) The reaction rate per $\alpha$ triplet according to Eq. (\ref{e3}) (top), the same quantity normalized by the corrected NACRE data (middle) and the temperature dependence (bottom). The original (\cite{Angulo1999}) and corrected NACRE data, as well as the H$\alpha$C and NM results are shown, according to the key. We also emphasize the astronomical constraints by Suda \textit{et. al} (\cite{Suda2011}) with purple points.}
\label{Fig3}
\end{figure}

%\maketitle
%%*******************************************************************
%%Introduction
%%*******************************************************************
In the hierarchy of stellar nucleosynthesis, carbon is produced in the helium burning stage. It takes place after the main sequence H-burning (\cite{Iliadis2007}), at temperatures of $T\sim 10^8$ K (\cite{rolfs1988cauldrons}) and  leads stars through the Red Giant and Asymptotic Giant Branch (AGB) phases (\cite{Hayashi1962}). It also plays a central role in more exotic phenomena, such as Type I supernovae in Neutron Stars and X-ray bursts in accreting White Dwarfs (\cite{Nomoto1985}). Its astrophysical significance has led to constraints of the reaction rates (\cite{Suda2011}), in order for results to be consistent with observations.\\
\indent The mechanism of He-burning corresponds to the fusion of 3 $\alpha$ particles and the subsequent decay of the excited ${}^{12}$C via sequential $\gamma$ emission (\cite{Saltpeter1952}). The abundance of carbon in the universe was explained in terms of a $0^+$ resonant state at $E_H=7.654$ MeV by Hoyle (\cite{Hoyle1954}) (the so-called ``Hoyle State"), which significantly accelerates the reaction, when compared to later burning stages, such as the $\alpha+{}^{12}\text{C}$ (\cite{deBoer2017,Plag,Nomoto1985,Langanke1986,Garrido2011,Ogata2010,Ishikawa2013,Angulo1999}).\\
\indent We propose a theoretical framework for fusion-evaporation reactions involving charged particles and/or photons. Our results and methodology are compared to the NACRE compilation (\cite{Angulo1999,NACREII2013}), also including our critical modifications to it.\\
%%*******************************************************************
%%Theory
%%*******************************************************************
\indent The triple alpha process is given by the following two steps (\cite{Hayashi1962}):
\begin{align*}
\alpha +\alpha & \rightarrow {}^{8}\textit{Be}^* \\
\alpha +{}^{8}\textit{Be}^* & \rightarrow {}^{12}\textit{C} + \gamma +\gamma '.
\label{echem}
\end{align*}\ignorespacesafterend
The first step, is the fusion product of two alpha particles but into beryllium-8, which is unstable and quickly decays back into its alpha constituents, with width $\Gamma_{\alpha,Be}^{(0)}=5.6$ eV (\cite{Angulo1999}). It is important to stress that for this process to occur the center of mass energy of the alphas must be larger than $E_{{}^{8}\textit{Be}}=-Q_1=92.08$ keV (\cite{Angulo1999}). This feature was not taken into account into the NACRE compilations and previous literature (\cite{Angulo1999,Ogata2010,Langanke1986,Ishikawa2013}), leading to a large overestimate of the reaction rates at the lowest temperatures, as shown below. Before beryllium decays, in the second step, carbon is formed predominantly via $\alpha$-fusion-$\gamma$-evaporation. It emits two $E2$ photons, one from the continuum above the alpha-threshold to the $2^+$, $4.43$ MeV state and one from that state to the $0^+$ ground state with very small cross sections (\cite{Hayashi1962}). This results in the $\alpha$ and $\gamma$ widths of the Hoyle state, i.e., $\Gamma_{\alpha,H}^{(0)}=9.3$ eV and $\Gamma_{\gamma,H}^{(0)}=5.1$ meV (\cite{BishopEfimov2021}). The total reaction rate $R_{3\alpha}$ is written as (\cite{Angulo1999}):
\begin{equationc}
R_{3\alpha}=\frac{\rho_{\alpha}^3}{3!} \langle \alpha\alpha\alpha\rangle
\label{e2}
\end{equationc}
where $\rho_{\alpha}$  is the number density of alphas and the reaction rate per $\alpha$-triplet $\langle \alpha\alpha\alpha\rangle$ is given by:
\begin{equationc}
N_A^2\langle \alpha\alpha\alpha\rangle = 
3 N_A \sqrt{\frac{8\pi}{\mu_{\alpha\alpha}\left(kT\right)^3}}\int_{E_{{}^{8}\textit{Be}}}^{E_B} \frac{\hbar}{\Gamma_{\alpha,1} }\sigma_1 \lbrace N_A \langle \sigma_2 v \rangle \rbrace e^{-E/kT}E dE
\label{e3}
\end{equationc}
with $\Gamma_{\alpha,1}$ and $\sigma_1=\sigma_{{}^{8}Be}$ being the width and cross section of the first step for center of mass energy $E$. The quantity $N_A\langle\sigma_2 v \rangle (E^*)$ is the reaction rate per $\alpha$ - beryllium pair for the second step, which depends on the excitation energy of beryllium, i.e., $E^*\equiv E>E_{{}^{8}\textit{Be}}$. This is given by (\cite{Angulo1999}):
\begin{equationc}
N_A\langle\sigma_2 v \rangle (E^*) = N_A \sqrt{\frac{8\pi}{\mu_{\alpha Be}\left(kT\right)^3}}\int_{0}^{E_B} \sigma_2 e^{-E'/kT} E' dE'
\label{e4}
\end{equationc}
where $\sigma_2=\sigma_{{}^{12}C} (E';E^*)$ is the cross section for the second step. This depends both on the excitation energy of beryllium $E^*$ and the center of mass energy of the $\alpha$ - beryllium pair $E'$, although the dependence on the former is weak, as we explain below. $N_A$ is the Avogadro's number. We stress that the integrals are calculated from the threshold energy ($E_{{}^{8}\textit{Be}}$ for step 1 and $0$ MeV for step 2) up to the energy of the barrier $E_B$. \\
\indent Since the cross sections are dominated by the $0^+$ resonances of beryllium and carbon in the energy regime of interest, we adopt a statistical description for the reactions (\cite{NACREII2013}). The reactant nuclei fuse creating a compound nucleus, which decays via an $\alpha$ or a $\gamma$ channel. The cross sections are written as:
\begin{equationc}
\sigma_{{}^{8}\textit{Be}} (E)= \sigma \left( \alpha\alpha, f\right)
\label{e5}
\end{equationc}
\begin{equationp}
\sigma_{{}^{12}\textit{C}} (E';E^*)= \sigma \left( \alpha{}^{8}\textit{Be}, f\right)\frac{\Gamma_{\gamma,2}}{\Gamma_2}
\label{e6}
\end{equationp}
The cross sections $\sigma \left( ..., f\right)$ describe the formation of the compound nucleus via fusion, while the branching ratio $\Gamma_{\gamma,2}/\Gamma_2$ of the widths ($\Gamma_{\gamma,2}$ gamma width and $\Gamma_2=\Gamma_{\alpha,2}+\Gamma_{\gamma,2}$ total width) for the second step, gives the probability of the $\gamma$ decay. The probability of the second $\gamma '$  decay is one, since it is the only available decay channel of the $2^+$ level of ${}^{12}$C.\\
\indent The fusion cross section are obtained via the use of two semi-classical models, coupled with the Feynman Path Integral method in Imaginary Time (ITM), described extensively in Refs. (\cite{BonaseraITM1994,BonaseraITM2000,BonaseraNatowitz2020,BonaseraEPJ2021,Depastas2023}). The key component of the method is the classical simulation of quantum tunneling via the evolution of the reaction system in imaginary time. This in effect reverses the sign of the collective forces below the barrier and thus, allows the system to tunnel between the classical turning points. Both the tunneling probability with relative angular momentum $l$, i.e., the penetrability $T_l\left(E_{CM}\right)$, as well as the imaginary time interval $\tau$, can be extracted. We note that in this study, we limit our calculations to $l=0$, due to its dominant contribution in the low energy region. The penetrability is calculated by the action of the system in imaginary time (\cite{BonaseraKimura2007,Depastas2023}), that is the integral of the imaginary momenta over the relative distance between the two turning points. The integrand values are obtained through semi-classical calculations with two different models, the Hybrid $\alpha$-Clustering model (H$\alpha$C) and the Neck Model (NM).\\
\indent The H$\alpha$C model (\cite{ZhengHac2021}) is a dynamical description of $A=4N_{\alpha}$ nuclei, with the $\alpha$ particles as the fundamental degrees of freedom. Their time evolution is governed by Hamiltonian equations of motion and their interaction contains a Bass nuclear potential (\cite{Bass1977}), a Coulomb potential and a ``Fermi interaction", that simulates the effect of Heisenberg and Pauli correlations of the constituents of the $\alpha$ particles. Due to the microscopic nature of the model, we are able to study reactions with excited species, but we are unable to consider the effect of resonances of the compound nucleus. On the contrary, the NM (\cite{BonaseraNM1984}) is a macroscopic description based on the Wigner transform of the Time-Dependent Hartree-Fock densities of the reactant nuclei. The nuclei interact via the Coulomb and Bass potentials, with the latter been modified to induce resonant structures, according to the prescription of Ref. (\cite{BonaseraEPJ2021}). The reason of using two models to calculate the fusion dynamics is their complementary characteristics. Both models give a reasonable description of fusion cross section below and above the Coulomb barrier (\cite{Depastas2023,BonaseraEPJ2021,BonaseraITM1994,BonaseraNM1984}).\\
\indent Other than the fusion cross sections, our theoretical framework requires the $\alpha$ and $\gamma$ widths, which can be written in terms of Fermi's Golden Rule:
\begin{equationc}
\Gamma_{\alpha/\gamma}=2\pi\left| M_{\alpha/\gamma} \right|^2 \rho(E_{\alpha/\gamma})
\label{e7}
\end{equationc}
with $M_{\alpha/\gamma}$ being the matrix element and $\rho$ the final state density of states. Following the Refs. (\cite{Gurtvitz1988,Buttike1982}), the width is:
\begin{equationc}
\Gamma_{\alpha}=\frac{\hbar}{\tau}T_0
\label{e8}
\end{equationc}
where $T_0=\left(1+e^{2A/\hbar} \right)^{-1}$ is the penetrability for $l=0$ as a function of the action $A$ in imaginary time (\cite{BonaseraITM1994}). No free parameters are needed in our calculations of the $\alpha$- widths as function of the energy.\\
\indent The $\gamma$ width is given by the standard description (\cite{ManyBodyProblem}):
\begin{equationc}
\Gamma_{\gamma}=\frac{8\pi(L+1)}{L[(2L+1)!!]^2}\left(\frac{E_{\gamma}}{\hbar c}\right)^{2L+1}B(E_I;L)
\label{e9}
\end{equationc}
where $L$ is the photon angular momentum (here $L=2$), $E_{\gamma}=E_I-E_F$ is the photon energy which equals the energy difference between the initial and final levels and $B(E_I;L)$ is the reduced matrix element. Using Eq. (\ref{e9}) at the resonance energy $E_I=E_R$, we get easily the general formula for the $\gamma$ decay width:
\begin{equationc}
\Gamma_{\gamma}=\Gamma_{\gamma}^{(0)}\left(\frac{E_I-E_F}{E_R-E_F}\right)^{2L+1}\frac{B(E_I;L)}{B(E_R;L)}
\label{e11}
\end{equationc}
with $\Gamma_{\gamma}^{(0)}$ being the $\gamma$ width at the resonance. In the literature (\cite{NACREII2013,BishopEfimov2021,Ogata2010,Angulo1999}) the last fraction in Eq. (\ref{e11}) is approximated to 1. This is a strong assumption implying that the matrix element is the same at the resonance (Hoyle state) and away from it. In contrast, we assume that the resonant behavior of the matrix element to be similar for both the $\alpha$ and $\gamma$ channels. Therefore, the reduced matrix element $B(E_I;L)$ is proportional to the $\alpha$ matrix element, times the density of states of the compound nucleus, that is $B(E_I;L) \sim \left| M_{\alpha} \right|^2 \rho(E_{\alpha}) \sim \Gamma_{\alpha}$ and Eq. (\ref{e11}) is rewritten as:
\begin{equationc}
\Gamma_{\gamma}=\Gamma_{\gamma}^{(0)}\left(\frac{E_I-E_F}{E_R-E_F}\right)^{2L+1}\frac{\Gamma_{\alpha}(E_I-Q_2)}{\Gamma_{\alpha}^{(0)}}
\label{e12}
\end{equationc}
where $Q_2=7.3667$ MeV (\cite{NNDC2022}) is the Q-value for the second step, $\Gamma_{\alpha}^{(0)}$ is the $\alpha$ width of the resonant state and $\Gamma_{\alpha}$ both given by Eq. (\ref{e8}).\\
%\begin{table}[b]%The best place to locate the table environment is directly after its first reference in text
%\caption{\label{tab1}%
%Experimental parameters of the 3$\alpha$ process, taken from Refs. \cite{NNDC2022,Angulo1999,BishopEfimov2021}. All energies are given in MeV.
%}
%\begin{ruledtabular}
%\begin{tabular}{lcccc}
%\textrm{Reaction}&
%\textrm{$Q$}&
%\textrm{$E_R$}&
%\textrm{$\Gamma_{\alpha}^{(0)}$}&
%\textrm{$\Gamma_{\gamma}^{(0)}$}\\
%\colrule
%$\alpha(\alpha,f){}^{8}\textit{Be}^*$ & -0.09208 & 0.09208 \footnote{$E_R=E_{{}^{8}\textit{Be}}$} & 5.6 10$^{-6}$\\
%${}^{8}\textit{Be}^*(\alpha,\gamma\gamma'){}^{12}\textit{C}$ & 7.3667 & 7.654 \footnote{$E_R=E_{H}$} & 9.3 10$^{-6}$ & 5.1 10$^{-9}$\\
%\end{tabular}
%\end{ruledtabular}
%\end{table}
\indent One of the most complete and widely used methodology in the literature for calculating the astrophysical reaction rates is that of the NACRE collaboration (\cite{Angulo1999,NACREII2013}). The reaction cross sections (denoted as $\sigma^N$) near the resonances are written in a Breit-Wigner form:
\begin{equationc}
\sigma^N = \frac{\pi}{k^2}\omega_l \frac{\Gamma_i\Gamma_f}{\left(E-E_R\right)^2+\left(\Gamma/2\right)^2}
\label{e13}
\end{equationc}
with $\omega_l$ a statistical factor including the spins and effects of identical nuclei, $k$ the wavenumber of the center of mass energy $E$ and relative angular momentum $l$, $\Gamma_i$, $\Gamma_f$ and $\Gamma$ the widths of the entrance, exit and total reaction, respectively. The cross section of the second step, is then given by:
\begin{multline}
\sigma_2^N = \frac{\pi}{k^2}\omega_l \frac{\Gamma_{\alpha,2}\Gamma_{\gamma,2}}{\left(E-E_R\right)^2+\left(\Gamma_2/2\right)^2}\approx\\\frac{\pi}{k^2}\omega_l \frac{\Gamma_{\alpha,2}^2}{\left(E-E_R\right)^2+\left(\Gamma_{\alpha,2}/2\right)^2}\frac{\Gamma_{\gamma,2}}{\Gamma_{\alpha,2}}=\sigma \left( \alpha{}^{8}\textit{Be}, f\right)\frac{\Gamma_{\gamma,2}}{\Gamma_{\alpha,2}}
\label{e14}
\end{multline}
\indent The difference between Eqs. (\ref{e6}) and (\ref{e14}) is striking. The denominator of the latter contains only the $\alpha$ width, instead of the total. For high energies, this is a more or less valid approximation, but in the limit of zero energy, the cross section is overestimated. This has important effects for the reaction rate at low temperatures, as we show later. The $\alpha$ and $\gamma$ widths are given by (\cite{Angulo1999,NACREII2013}):
\begin{equationc}
\Gamma_{\alpha}=\Gamma_{\alpha}^{(0)}\frac{T_l(E)}{T_l(E_R)}
\label{e15}
\end{equationc}
\begin{equationp}
\Gamma_{\gamma}=\Gamma_{\gamma}^{(0)}\left(\frac{E_I-E_F}{E_R-E_F}\right)^{2L+1}
\label{e16}
\end{equationp}
\indent In the case of the $\alpha$ width, in their approach it is proportional to the penetrability and normalized to the resonant $\Gamma_{\alpha}^{(0)}$ value, similarly to Ref. (\cite{Lane1958}). In a similar manner, the $\gamma$ width retains its proportionality to the fifth power of the energy and the normalization to the resonant $\Gamma_{\gamma}^{(0)}$. An important difference is that in the NACRE methodology, the integrals of Eqs. (\ref{e3}) and (\ref{e4}) are calculated from $0$ MeV, which is not correct, because of the threshold value for ${}^{8}$Be-fusion, i.e., step 1.\\
%%*******************************************************************
%%Results
%%*******************************************************************
\indent In Fig. \ref{Fig1} we plot the fusion cross sections (left) and $\alpha$ widths as functions of $E_{CM}$ (right). The NACRE cross sections are smaller than the H$\alpha$C and NM (apart at the resonance). We also notice that our calculated cross sections correctly start from $E_{{}^{8}\textit{Be}}$ while the NACRE results do not.\\
\indent The calculated widths (NM) are similar with the corresponding experimental values (\cite{Angulo1999,BishopEfimov2021}) (black points). Also, we plot the widths with the NACRE formulas Eq. (\ref{e16}) using two penetrability formulas, the one from the ITM which accounts for both the Coulomb and Nuclear potentials (see Ref. \cite{Depastas2023}) and the other from the quantum mechanical Coulomb scattering (see Ref. \cite{NACREII2013}). We stress that the widths calculated from Eq. (\ref{e9}), are resulting purely by the sub-barrier nuclear dynamics in imaginary times. On the contrary, in Ref. (\cite{Angulo1999}) the normalization to the resonant value, results in much higher widths for the entire energy region.\\
%***************************************************************************
%%% Figure
%***************************************************************************
\indent We present the results of the $(\alpha,\gamma\gamma')$ cross sections for different excitation energies in Fig. \ref{Fig2}. On the top panel the data for the fusion of the beryllium ground state are shown, while on the bottom panel the results for $E^*=0.10$, $0.50$ and $0.75$ MeV are given. The NACRE cross sections are smaller, while the $\Gamma_{\gamma}/\Gamma_{\alpha}$ and $\Gamma_{\gamma}/\Gamma$ formulas Eqs. (\ref{e6}) and (\ref{e14}) differ only in the low energy region, as expected. The NM and H$\alpha$C results are similar, apart from the resonance not included in the HaC model. The $\gamma$ widths given by Eq. (\ref{e16}), are normalized to their resonant values, which leads to higher values of $\Gamma_{\gamma}$ for the NACRE data.\\
\indent On the bottom panel, we see that the cross sections of the H$\alpha$C model are almost independent of the excitation energy of the beryllium. This allows us to factor out the $N_A \langle \sigma_2 v \rangle$ rate from the integral in Eq. (\ref{e3}). In the case of the H$\alpha$C model, we take this to be equal to an average rate plus an error, both resulting from the data of Fig. \ref{Fig2}, i.e.,  $N_A \langle \sigma_2 v \rangle (E^*) \rightarrow N_A \left( \overline{\langle \sigma_2 v \rangle} \pm \delta \langle \sigma_2 v \rangle \right)$. For the NM, we use the ground state rate, i.e., $N_A \langle \sigma_2 v \rangle (E^*) \rightarrow N_A \langle \sigma_2 v \rangle (E^*=0\text{ MeV})$.\\
%***************************************************************************
%%% Figure
%***************************************************************************
\indent Finally, we present the stellar reaction rates according to Eq. (\ref{e3}) in Fig. \ref{Fig3} (top). There we show the H$\alpha$C (red) and NM (blue) calculations, the original NACRE (green) (\cite{Angulo1999}), as well as the NACRE with the corrections of widths and threshold energies (black). These corrections consist of using the width ratio $\Gamma_{\gamma,2}/\Gamma_{2}$ instead of $\Gamma_{\gamma,2}/\Gamma_{\alpha,2}$ in Eq. (\ref{e14}) and integrating the reaction rates from the threshold energy $E_{{}^{8}\textit{Be}}$ instead from $0$ MeV, similarly to Eq. (\ref{e3}). In the middle panel, we re-plot the same quantities normalized by the corresponding corrected NACRE values. On the bottom panel, we illustrate the temperature dependence $\nu$ of the reaction rate, defined in the figure.\\
%***************************************************************************
%%% Figure
%***************************************************************************
\indent We observe that the original NACRE prescription has an enhanced rate by almost 20-30 orders of magnitude for temperatures $T\sim 10^7$ K, when compared to the other approaches (\cite{Ogata2010,Ishikawa2013,Nguyen2013,Garrido2011}). This may also have significant consequences for stellar dynamics, since the helium burning may start prematurely in stars of lower temperature than $T\sim 10^{7.6}\text{-}10^8$ K (\cite{Suda2011}), as discussed in the literature. We emphasize that the H$\alpha$C results are lower than the other approaches, due to the lack of resonant peaks, while the NM data are higher than the corrected NACRE and the H$\alpha$C, due to the existence of such structures and the increased $\gamma$ cross section (as shown in Fig. \ref{Fig2}).\\
\indent In the top panel of Fig. 3, we also include an astrophysical upper limit by Suda \textit{et. al} (\cite{Suda2011}). This limit restricts the maximum reaction rate at $T\sim 10^{7.8}$ K to the value of $10^{-29}$ cm$^{6}$s$^{-1}$mol$^{-2}$. The constraint comes from the requirement that the helium ignition takes place at high enough temperature in order for the First Asymptotic Giant Branch (FBG) stars to have a value of luminosity in agreement to observations.\\
\indent Apart from the upper rate constraint, Suda \textit{et. al} (\cite{Suda2011}) proposed a lower limit in the temperature dependence $\nu$. Specifically, $\nu\gtrsim 10$ in the region $T\sim 10^8-1.2 \cdot 10^8$ K. This requirement ensures that Asymptotic Giant Branch (AGB) stars experience helium flashes on their shells, which result is third dredge-up events that explain the presence of S-process elements such as ${}^{99}$Tc in their outer envelopes (\cite{Suda2011}). Thus, more stringent constraints are required for the helium burning rates and these might come from Nuclear Physics. Studies could be performed through the utilization of the Trojan Horse Method (THM) (\cite{Tumino2018}), which is able to effectively bypass the issue of the small life time of beryllium-8 and the suppression of the cross section due to Coulomb interaction, via the use of a heavier clusterized nuclear system (see Fig. \ref{Fig2}, top).\\
%%*******************************************************************
%%Conclusion
%%*******************************************************************
\indent To summarize, we developed a theoretical framework for the low energy astrophysical fusion-evaporations involving charged particles and/or photons based on the statistical breakup of a compound nucleus, formed through the tunneling of the reactant system. The fusion cross sections are calculated via the Feynman Path Integral method in Imaginary Time, coupled with the semi-classical H$\alpha$C and NM models. We furthermore derive approximate formulas for the charged particle and photon widths.\\
\indent For the 3$\alpha$ process and we find results consistent with both nuclear and astrophysical constraints. Finally, we note the need for stringer constraints of the functional form of cross sections and we propose possible experimental investigations with the Trojan Horse Method.\\
\\ACKNOWLEDGEMENTS\\
This work work was supported in part by the United States Department of Energy under Grant $\#$DE-FG03-93ER40773, NNSA Grant No. DENA0003841 (CENTAUR) and by the National
Natural Science Foundation of China (Grant Nos. 11905120 and 11947416).
%***************************************************************************
%%% References
%***************************************************************************
\bibliographystyle{elsarticle-harv} 
\bibliography{references}{}

\providecommand{\noopsort}[1]{}\providecommand{\singleletter}[1]{#1}%
\begin{thebibliography}{32}
\expandafter\ifx\csname natexlab\endcsname\relax\def\natexlab#1{#1}\fi
\providecommand{\url}[1]{\texttt{#1}}
\providecommand{\href}[2]{#2}
\providecommand{\path}[1]{#1}
\providecommand{\DOIprefix}{doi:}
\providecommand{\ArXivprefix}{arXiv:}
\providecommand{\URLprefix}{URL: }
\providecommand{\Pubmedprefix}{pmid:}
\providecommand{\doi}[1]{\href{http://dx.doi.org/#1}{\path{#1}}}
\providecommand{\Pubmed}[1]{\href{pmid:#1}{\path{#1}}}
\providecommand{\bibinfo}[2]{#2}
\ifx\xfnm\relax \def\xfnm[#1]{\unskip,\space#1}\fi
%Type = Article
\bibitem[{Angulo et~al.(1999)Angulo, Arnould, Rayet, Descouvemont, Baye, Leclercq-Willain, Coc, Barhoumi, Aguer, Rolfs, Kunz, Hammer, Mayer, Paradellis, Kossionides, Chronidou, Spyrou, Degl'Innocenti, Fiorentini, Ricci, Zavatarelli, Providencia, Wolters, Soares, Grama, Rahighi, Shotter and {Lamehi Rachti}}]{Angulo1999}
\bibinfo{author}{Angulo, C.}, \bibinfo{author}{Arnould, M.}, \bibinfo{author}{Rayet, M.}, \bibinfo{author}{Descouvemont, P.}, \bibinfo{author}{Baye, D.}, \bibinfo{author}{Leclercq-Willain, C.}, \bibinfo{author}{Coc, A.}, \bibinfo{author}{Barhoumi, S.}, \bibinfo{author}{Aguer, P.}, \bibinfo{author}{Rolfs, C.}, \bibinfo{author}{Kunz, R.}, \bibinfo{author}{Hammer, J.}, \bibinfo{author}{Mayer, A.}, \bibinfo{author}{Paradellis, T.}, \bibinfo{author}{Kossionides, S.}, \bibinfo{author}{Chronidou, C.}, \bibinfo{author}{Spyrou, K.}, \bibinfo{author}{Degl'Innocenti, S.}, \bibinfo{author}{Fiorentini, G.}, \bibinfo{author}{Ricci, B.}, \bibinfo{author}{Zavatarelli, S.}, \bibinfo{author}{Providencia, C.}, \bibinfo{author}{Wolters, H.}, \bibinfo{author}{Soares, J.}, \bibinfo{author}{Grama, C.}, \bibinfo{author}{Rahighi, J.}, \bibinfo{author}{Shotter, A.}, \bibinfo{author}{{Lamehi Rachti}, M.}, \bibinfo{year}{1999}.
\newblock \bibinfo{journal}{Nuclear Physics A} \bibinfo{volume}{656}, \bibinfo{pages}{3--183}.
%Type = Article
\bibitem[{Bass(1977)}]{Bass1977}
\bibinfo{author}{Bass, R.}, \bibinfo{year}{1977}.
\newblock \bibinfo{journal}{Phys. Rev. Lett.} \bibinfo{volume}{39}, \bibinfo{pages}{265}.
%Type = Article
\bibitem[{Basunia and Chakraborty(2022)}]{NNDC2022}
\bibinfo{author}{Basunia, M.}, \bibinfo{author}{Chakraborty, A.}, \bibinfo{year}{2022}.
\newblock \bibinfo{journal}{Nucl. Data Sheets} \bibinfo{volume}{186}.
%Type = Article
\bibitem[{Bemmerer and et~al.(2011)}]{Garrido2011}
\bibinfo{author}{Bemmerer, E.}, \bibinfo{author}{et~al.}, \bibinfo{year}{2011}.
\newblock \bibinfo{journal}{Eur. Phys. J. A} \bibinfo{volume}{47}, \bibinfo{pages}{102}.
%Type = Article
\bibitem[{Bishop et~al.(2021)Bishop, Rogachev, Ahn, Aboud, Barbui, Bosh, Hooker, Hunt, Jayatissa, Koshchiy, Malecek, Marley, Munch, Pollacco, Pruitt, Roeder, Saastamoinen, Sobotka and Upadhyayula}]{BishopEfimov2021}
\bibinfo{author}{Bishop, J.}, \bibinfo{author}{Rogachev, G.V.}, \bibinfo{author}{Ahn, S.}, \bibinfo{author}{Aboud, E.}, \bibinfo{author}{Barbui, M.}, \bibinfo{author}{Bosh, A.}, \bibinfo{author}{Hooker, J.}, \bibinfo{author}{Hunt, C.}, \bibinfo{author}{Jayatissa, H.}, \bibinfo{author}{Koshchiy, E.}, \bibinfo{author}{Malecek, R.}, \bibinfo{author}{Marley, S.T.}, \bibinfo{author}{Munch, M.}, \bibinfo{author}{Pollacco, E.C.}, \bibinfo{author}{Pruitt, C.D.}, \bibinfo{author}{Roeder, B.T.}, \bibinfo{author}{Saastamoinen, A.}, \bibinfo{author}{Sobotka, L.G.}, \bibinfo{author}{Upadhyayula, S.}, \bibinfo{year}{2021}.
\newblock \bibinfo{journal}{Phys. Rev. C} \bibinfo{volume}{103}, \bibinfo{pages}{L051303}.
%Type = Article
\bibitem[{Bonasera(2021)}]{BonaseraEPJ2021}
\bibinfo{author}{Bonasera, A.}, \bibinfo{year}{2021}.
\newblock \bibinfo{journal}{EPJ Web Conf.} \bibinfo{volume}{252}, \bibinfo{pages}{05001}.
%Type = Article
\bibitem[{Bonasera et~al.(1984)Bonasera, Bertsch and El-Sayed}]{BonaseraNM1984}
\bibinfo{author}{Bonasera, A.}, \bibinfo{author}{Bertsch, G.}, \bibinfo{author}{El-Sayed, E.}, \bibinfo{year}{1984}.
\newblock \bibinfo{journal}{Phys. Lett. B} \bibinfo{volume}{141}.
%Type = Article
\bibitem[{Bonasera and Kondratyev(1994)}]{BonaseraITM1994}
\bibinfo{author}{Bonasera, A.}, \bibinfo{author}{Kondratyev, V.}, \bibinfo{year}{1994}.
\newblock \bibinfo{journal}{Phys. Lett. B} \bibinfo{volume}{339}, \bibinfo{pages}{207--210}.
%Type = Article
\bibitem[{Bonasera and Natowitz(2020)}]{BonaseraNatowitz2020}
\bibinfo{author}{Bonasera, A.}, \bibinfo{author}{Natowitz, J.B.}, \bibinfo{year}{2020}.
\newblock \bibinfo{journal}{Phys. Rev. C} \bibinfo{volume}{102}, \bibinfo{pages}{061602}.
%Type = Article
\bibitem[{B\"uttiker and Landauer(1982)}]{Buttike1982}
\bibinfo{author}{B\"uttiker, M.}, \bibinfo{author}{Landauer, R.}, \bibinfo{year}{1982}.
\newblock \bibinfo{journal}{Phys. Rev. Lett.} \bibinfo{volume}{49}, \bibinfo{pages}{1739--1742}.
%Type = Article
\bibitem[{deBoer et~al.(2017)deBoer, G\"orres, Wiescher, R.E. Azumaand~Best, Brune, Fields, Jones, Pignatari, Sayre, Smith, Timmes and Uberseder}]{deBoer2017}
\bibinfo{author}{deBoer, R.J.}, \bibinfo{author}{G\"orres, J.}, \bibinfo{author}{Wiescher, M.}, \bibinfo{author}{R.E. Azumaand~Best, A.}, \bibinfo{author}{Brune, C.R.}, \bibinfo{author}{Fields, C.E.}, \bibinfo{author}{Jones, S.}, \bibinfo{author}{Pignatari, M.}, \bibinfo{author}{Sayre, D.}, \bibinfo{author}{Smith, K.}, \bibinfo{author}{Timmes, F.X.}, \bibinfo{author}{Uberseder, E.}, \bibinfo{year}{2017}.
\newblock \bibinfo{journal}{Rev. Mod. Phys.} \bibinfo{volume}{89}, \bibinfo{pages}{035007}.
%Type = Article
\bibitem[{Depastas et~al.(2023)Depastas, Sun, Zheng and Bonasera}]{Depastas2023}
\bibinfo{author}{Depastas, T.}, \bibinfo{author}{Sun, S.T.}, \bibinfo{author}{Zheng, H.}, \bibinfo{author}{Bonasera, A.}, \bibinfo{year}{2023}.
\newblock \bibinfo{journal}{Phys. Rev. C} \bibinfo{volume}{108}, \bibinfo{pages}{035806}.
%Type = Article
\bibitem[{Gurvitz(1988)}]{Gurtvitz1988}
\bibinfo{author}{Gurvitz, S.}, \bibinfo{year}{1988}.
\newblock \bibinfo{journal}{Phys. Rev. A} \bibinfo{volume}{38}, \bibinfo{pages}{1747--1759}.
%Type = Article
\bibitem[{Hayashi et~al.(1962)Hayashi, {H{\={o}}shi} and {Sugimoto}}]{Hayashi1962}
\bibinfo{author}{Hayashi, C.}, \bibinfo{author}{{H{\={o}}shi}, R.}, \bibinfo{author}{{Sugimoto}, D.}, \bibinfo{year}{1962}.
\newblock \bibinfo{journal}{Progress of Theoretical Physics Supplement} \bibinfo{volume}{22}, \bibinfo{pages}{1--183}.
%Type = Article
\bibitem[{Hoyle(1954)}]{Hoyle1954}
\bibinfo{author}{Hoyle, F.}, \bibinfo{year}{1954}.
\newblock \bibinfo{journal}{Astrophys. J. Suppl.} \bibinfo{volume}{1}, \bibinfo{pages}{121}.
%Type = Book
\bibitem[{{Iliadis}(2007)}]{Iliadis2007}
\bibinfo{author}{{Iliadis}, C.}, \bibinfo{year}{2007}.
\newblock \bibinfo{title}{{Nuclear Physics of Stars}}.
\newblock \bibinfo{publisher}{Wiley, Weinheim}.
%Type = Article
\bibitem[{Ishikawa(2013)}]{Ishikawa2013}
\bibinfo{author}{Ishikawa, S.}, \bibinfo{year}{2013}.
\newblock \bibinfo{journal}{Phys. Rev. C} \bibinfo{volume}{87}, \bibinfo{pages}{055804}.
%Type = Article
\bibitem[{Kimura and Bonasera(2007)}]{BonaseraKimura2007}
\bibinfo{author}{Kimura, S.}, \bibinfo{author}{Bonasera, A.}, \bibinfo{year}{2007}.
\newblock \bibinfo{journal}{Phys. Rev. C} \bibinfo{volume}{76}, \bibinfo{pages}{031602}.
%Type = Article
\bibitem[{Kondratyev et~al.(2000)Kondratyev, Bonasera and Iwamoto}]{BonaseraITM2000}
\bibinfo{author}{Kondratyev, V.}, \bibinfo{author}{Bonasera, A.}, \bibinfo{author}{Iwamoto, A.}, \bibinfo{year}{2000}.
\newblock \bibinfo{journal}{Phys. Rev. C} \bibinfo{volume}{61}, \bibinfo{pages}{044613}.
%Type = Article
\bibitem[{Lane and Thomas(1958)}]{Lane1958}
\bibinfo{author}{Lane, A.M.}, \bibinfo{author}{Thomas, R.}, \bibinfo{year}{1958}.
\newblock \bibinfo{journal}{Rev. Mod. Phys.} \bibinfo{volume}{30}, \bibinfo{pages}{257--353}.
%Type = Article
\bibitem[{Langanke et~al.(1986)Langanke, Wiescher and Thielemann}]{Langanke1986}
\bibinfo{author}{Langanke, K.}, \bibinfo{author}{Wiescher, M.}, \bibinfo{author}{Thielemann, F.}, \bibinfo{year}{1986}.
\newblock \bibinfo{journal}{Z. Physik A - Atomic Nuclei} \bibinfo{volume}{324}, \bibinfo{pages}{147–152}.
%Type = Article
\bibitem[{Nguyen et~al.(2013)Nguyen, Nunes and Thompson}]{Nguyen2013}
\bibinfo{author}{Nguyen, N.}, \bibinfo{author}{Nunes, F.}, \bibinfo{author}{Thompson, I.}, \bibinfo{year}{2013}.
\newblock \bibinfo{journal}{arXiv:1209.4999} .
%Type = Article
\bibitem[{Nomoto et~al.(1985)Nomoto, Thielemann and Miyaji}]{Nomoto1985}
\bibinfo{author}{Nomoto, K.}, \bibinfo{author}{Thielemann, F.K.}, \bibinfo{author}{Miyaji, S.}, \bibinfo{year}{1985}.
\newblock \bibinfo{journal}{Astron. Astrophys.} \bibinfo{volume}{149}, \bibinfo{pages}{239--245}.
%Type = Article
\bibitem[{Ogata et~al.(2010)Ogata, Kan and Kamimura}]{Ogata2010}
\bibinfo{author}{Ogata, K.}, \bibinfo{author}{Kan, M.}, \bibinfo{author}{Kamimura, M.}, \bibinfo{year}{2010}.
\newblock \bibinfo{journal}{AIP Conference Proceedings} \bibinfo{volume}{1269}, \bibinfo{pages}{268--275}.
%Type = Article
\bibitem[{Plag and et~al.(2012)}]{Plag}
\bibinfo{author}{Plag, R.}, \bibinfo{author}{et~al.}, \bibinfo{year}{2012}.
\newblock \bibinfo{journal}{Phys. Rev. C} \bibinfo{volume}{86}, \bibinfo{pages}{015805}.
%Type = Book
\bibitem[{Ring and Schuck(1980)}]{ManyBodyProblem}
\bibinfo{author}{Ring, P.}, \bibinfo{author}{Schuck, P.}, \bibinfo{year}{1980}.
\newblock \bibinfo{title}{The nuclear many-body problem}.
\newblock \bibinfo{publisher}{Springer-Verlag}, \bibinfo{address}{New York}.
%Type = Book
\bibitem[{Rolfs and Rodney(1988)}]{rolfs1988cauldrons}
\bibinfo{author}{Rolfs, C.E.}, \bibinfo{author}{Rodney, W.S.}, \bibinfo{year}{1988}.
\newblock \bibinfo{title}{{Cauldrons in the cosmos: Nuclear astrophysics}}.
\newblock \bibinfo{publisher}{University of Chicago press}.
%Type = Article
\bibitem[{Salpeter(1952)}]{Saltpeter1952}
\bibinfo{author}{Salpeter, E.E.}, \bibinfo{year}{1952}.
\newblock \bibinfo{journal}{Astrophys. J.} \bibinfo{volume}{115}, \bibinfo{pages}{326--328}.
%Type = Article
\bibitem[{Suda et~al.(2011)Suda, Hirschi and Fujimoto}]{Suda2011}
\bibinfo{author}{Suda, T.}, \bibinfo{author}{Hirschi, R.}, \bibinfo{author}{Fujimoto, M.}, \bibinfo{year}{2011}.
\newblock \bibinfo{journal}{The Astrophysical Journal} \bibinfo{volume}{741}, \bibinfo{pages}{61}.
%Type = Article
\bibitem[{Tumino et~al.(2018)Tumino, Spitaleri, Cognata, Cherubini, Guardo, Gulino, S.~Hayakawa, Lamia, Petrascu, Pizzone, Puglia, Rapisarda, Romano, Sergi, Spartá and Trache}]{Tumino2018}
\bibinfo{author}{Tumino, A.}, \bibinfo{author}{Spitaleri, C.}, \bibinfo{author}{Cognata, M.L.}, \bibinfo{author}{Cherubini, S.}, \bibinfo{author}{Guardo, G.L.}, \bibinfo{author}{Gulino, M.}, \bibinfo{author}{S.~Hayakawa, I.I.}, \bibinfo{author}{Lamia, L.}, \bibinfo{author}{Petrascu, H.}, \bibinfo{author}{Pizzone, R.G.}, \bibinfo{author}{Puglia, S.M.R.}, \bibinfo{author}{Rapisarda, G.G.}, \bibinfo{author}{Romano, S.}, \bibinfo{author}{Sergi, M.L.}, \bibinfo{author}{Spartá, R.}, \bibinfo{author}{Trache, L.}, \bibinfo{year}{2018}.
\newblock \bibinfo{journal}{Nature} \bibinfo{volume}{557}, \bibinfo{pages}{687–690}.
%Type = Article
\bibitem[{Xu et~al.(2013)Xu, Takahashi, Goriely, Arnould, Ohta and Utsunomiya}]{NACREII2013}
\bibinfo{author}{Xu, Y.}, \bibinfo{author}{Takahashi, K.}, \bibinfo{author}{Goriely, S.}, \bibinfo{author}{Arnould, M.}, \bibinfo{author}{Ohta, M.}, \bibinfo{author}{Utsunomiya, H.}, \bibinfo{year}{2013}.
\newblock \bibinfo{journal}{Nuclear Physics A} \bibinfo{volume}{918}, \bibinfo{pages}{61--169}.
%Type = Article
\bibitem[{Zheng and Bonasera(2021)}]{ZhengHac2021}
\bibinfo{author}{Zheng, H.}, \bibinfo{author}{Bonasera, A.}, \bibinfo{year}{2021}.
\newblock \bibinfo{journal}{Symmetry} \bibinfo{volume}{13}, \bibinfo{pages}{1777}.

\end{thebibliography}
\end{document}